\documentclass[aps,prb,twocolumn,amsmath,groupedaddress]{revtex4}  
\usepackage{graphicx}  
\usepackage{dcolumn}   
\usepackage{bm}        
\usepackage{amssymb}   
\usepackage{multirow}

\begin{document}

\title{Edge spin excitations and reconstructions of integer quantum Hall liquids}

\author{Yuhui Zhang, Kun Yang}
\affiliation{National High Magnetic Field Laboratory and
  Department of Physics, Florida State University, Tallahassee, FL
  32306, USA}

\date{\today}

\begin{abstract}
We study the effect of electron-electron interaction on the charge and spin structures at the edge of integer quantum Hall liquids, under three different kinds of confining potentials. Our exact diagonalization calculation for small systems indicates that the low energy excitations of $\nu=1$ ferromagnetic state are bosonic edge spin waves. Instabilities of the ferromagnetic state with altering confinement strength result from the softening of these edge spin waves, and formation of edge spin textures. In $\nu\lesssim 2$ regime, exact diagonalization on edge electron systems indicates that compact Hartree-Fock states with different total spin always become ground states in some regions of parameter space, and the ground states appear in between two compact states are their edge spin waves. The initial $\nu=2$ instability is toward the compact state with total spin $1$. Larger systems are studied using a microscopic trial wave functions, and some quantitative predictions on the edge instabilities for a certain type of confining potential are reached in the thermodynamic limit.

\end{abstract}

\maketitle

\section{INTRODUCTION}
In quantum Hall (QH) systems the charged bulk excitations are gapped, while gapless excitations exist at the edges.\cite{gapless_edge, Wen} Due to the quantization of kinetic energy under strong magnetic field, the physics of edge excitations depends on the interplay of electron-electron interaction and the confining potential. The different manners in which the two dimensional electron gas (2DEG) is confined lead to rich electronic structures at the edges of the sample.

The edge of a $\nu=1$ spinless QH liquid confined in a simple geometry by a sharp confining potential is described by the $Z_{F}=1$ chiral Fermi-liquid theory.\cite{gapless_edge} With smoother confinement, the edge undergoes a {\em charge reconstruction} transition in which separated electronic lumps appear and bring new boundaries with even number, breaking the chirality.\cite{EricYang1,Chamon} Further including the spin freedom and with sufficiently weak Zeeman splitting, Hartree-Fock (HF) technique shows that the charge reconstruction transition is pre-emptied by formation of edge spin texture.\cite{Sondhi96} In fact closely related edge {\em spin reconstruction} was found for $\nu=2$ within HF approximation, in which a spin polarized strip is created along the edge and through which the edge undergoes a second-order spin-unpolarized to spin-polarized transition as the confining potential smoothes.\cite{Dempsey, Yacoby} The edge modes associated with this spin reconstruction of $\nu=2$ edge show up in NMR\cite{NMR} and momentum-resolved electron tunneling spectroscopy.\cite{Yacoby}
Still within HF approximation, Barlas {\em et al.}\cite{Yafis} found that under sufficiently weak triangular confining potential, unpolarized $\nu=2$ state's spin-polarization transition is preceded by a charge reconstruction of a single spin species, creating a detached spin-polarized strip. This kind of edge structure are understood as a simultaneous edge {\em charge and spin reconstruction}. It is these recent theoretical and in particular, experimental developments that motivated the present work, which go beyond earlier theoretical work\cite{Sondhi96, Dempsey, Yafis, Sondhi99, Oaknin, Others} in several aspects.

First of all, earlier studies of integer QH edge instabilities and reconstructions were mostly based on HF approximation, and focused on the ground states only. In the present work we perform detailed exact diagonalization study on the same systems. In addition to obtaining the exact ground states, we also obtain the low-energy spectra of the system before instabilities occur, which provide deeper insight into the nature of the instabilities and the mechanisms of the corresponding edge reconstructions.

Secondly, details of a QH liquid's edge structure may also depend on the detailed form of the confining potential. Earlier works usually focus on one specific type of confining potential. In this paper, we consider a 2DEG with disk geometry under three different types of confinement. Positive background charge confining potential is usually used to model a realistic confinement.\cite{Wan}
Parabolic confining potential is commonly used in the research of QH dots.\cite{EricYang1,EricYang2} By numerically solving the Poisson and Schrodinger equations for electron states in a $GaAs/Al_{x}Ga_{1-x}As$ heterostructure with three spatial dimensions' confinement in Hartree approximation, Kumar {\em et al.} argue that it is reasonable to model a real QH dot device's confining potential as a parabolic one.\cite{Kumar}
Linear confining potential also deserves some attention, because the recent momentum-resolved tunneling experiments\cite{Yacoby} for parallel wires which can be modeled as linear confining potential, have observed edge spin reconstruction, and more complicated edge charge and spin reconstruction is found in a similar triangular confining potential.\cite{Yafis} Such a comprehensive study allows us to compare the systems' behavior under different types of confinements, and distinguish between universal and non-universal aspects of edge physics.

Last but not least, here we study $\nu=1$ and $\nu=2$ edges together, and reveal the similarities and differences in their behavior. Such side-by-side comparison has not been performed before. In fact insights from the study of $\nu=1$ are crucial to our understanding of the $\nu=2$ case.            

Our most robust results are summarized as what follows.
Ignoring Zeeman coupling, the low energy spectra of $\nu=1$ ferromagnetic state can be mapped onto those of $\Delta S=-1$ bosons on top of the ferromagnetic state. These excited bosons are edge spin waves (ESWs), which propagate in {\em both} directions.
Under positive background charge confining potential, the ferromagnetic state is stable as the distance $d$ from positive background charge layer to 2DEG approaches zero. When the distance $d$ becomes larger, ferromagnetic state is destabilized by softening of ESWs, resulting in edge spin textures. Under parabolic or linear confinement, the ferromagnetic state is destabilized by softening of ESWs under both smoother and stronger confinements, and the ferromagnetic state window disappears as the particle number $N$ increases.
In $\nu\lesssim 2$ regime, compact states (defined below) with different total spin $S$ can always become ground states in some regions of parameter space. The low energy excitations of a compact state can be mapped onto those of $\Delta S=-1$ bosons on top of it, thus are ESWs of this compact state. The ground states appearing in between two different compact states can thus be viewed as edge spin textures as well. The initial $\nu=2$ instability is toward the $S=1$ compact state.

The rest of the paper is organized as what follows. In Sec. II, we introduce the models, provide some numerical details of our calculations, and an electrostatic model which gives some qualitative understanding on integer quantum Hall liquid's edge instabilities. In Sec. III we investigate the low energy excitations and edge instabilities of $\nu=1$ ferromagnetic state using exact diagonalization on small systems and a microscopic trial wave function on large systems.
Sec. IV considers the low energy excitations and edge instabilities of $\nu=2$ unpolarized state. Besides carrying out exact diagonalization on small systems, we also study the large systems by separating out the edge electron systems and using exact diagonalization on them. Some concluding remarks are made in Sec. V. We make comparisons with earlier work wherever appropriate.

\section{THE MODELS AND ELECTROSTATIC CONSIDERATION}
\subsection{The models}
We consider a 2DEG with spin degree of freedom in a disk geometry. The single particle Hamiltonian is
\begin{equation}
\mathcal{H}=\frac{1}{2m_{e}}[\mathbf{p}+\frac{e}{c}\mathbf{A}(\mathbf{r})]^{2}+U(\mathbf{r})+g\mu_{B}s_{z}B   ,
\label{singleH}
\end{equation}
where $m_{e}$, $-e$ is the mass and charge of a single electron respectively. $c$ is the speed of light. $\mu_{B}$ is the Bohr magneton. $\mathbf{p}$ is the momentum operator. $U(\mathbf{r})$ is the rotationally invariant confining potential. The vector potential $\mathbf{A}=(-By/2,Bx/2,0)$ in symmetric gauge. $s_{z}$ is the $z$ axis component of single particle spin operator. Confining electrons to the lowest Landau level (LL), the single particle wave functions are
\begin{equation}
\phi_{m}(z)=(2\pi l_{B}^{2} 2^{m} m!)^{-1/2} z^{m} e^{-|z|^{2}/4},
\label{singleWf}
\end{equation}
where $z=(x-iy)/l_{B}$ is the complex coordinate in the plane of the 2DEG and $l_{B}=\sqrt{\hbar c/eB}$ is magnetic length, $m=0$, $1$, $2$, $\cdots$.
The complete Hamiltonian in symmetric gauge is then
\begin{equation}
\begin{split}
H =& \frac{1}{2}\underset{m,n,l,\sigma,\sigma'}{\sum}V_{mn}^{l}c_{m+l,\sigma}^{\dagger}c_{n,\sigma'}^{\dagger}c_{n+l,\sigma'}c_{m,\sigma}+\underset{m,\sigma}{\sum}U_{m}^{cp}\hat{n}_{m,\sigma} \\
&+\frac{1}{2}g\mu_{B}B\underset{m}{\sum}(\hat{n}_{m,\uparrow}-\hat{n}_{m,\downarrow}) ,
\end{split}
\label{completeH}
\end{equation}
where $c_{m,\sigma}^{\dagger}$ is the electron creation operator for the lowest LL single-electron state with orbital angular momentum $m$ and spin $\sigma$, $\hat{n}_{m,\sigma}=c_{m,\sigma}^{\dagger}c_{m,\sigma} $ is the occupation number operator of the $m$th orbital with spin $\sigma$, and
\small
\begin{equation}
V_{mn}^{l}=\int d^{2}r_{1}\int d^{2}r_{2}\phi_{m+l}^{*}(\mathbf{r_{1}})\phi_{n}^{*}(\mathbf{r_{2}})\frac{e^{2}}{\epsilon r_{12}}\phi_{n+l}(\mathbf{r_{2}})\phi_{m}(\mathbf{r_{1}})
\label{CoulombInt}
\end{equation}
\normalsize
represents the electron-electron Coulomb interaction ($\epsilon$ is the dielectric constant). $U_{m}^{cp}$ is the matrix element of the rotationally invariant confining potential.
Hamiltonian (\ref{completeH}) has rotational symmetry, and also commutes with total spin operator $\mathbf{S} $ and its $z$ axis component $S_{z}$. Total angular momentum $M$, total spin $S$ and $S_{z}$ are good quantum numbers.

In GaAs, arising from band structure and spin-orbit coupling, electron spin $g$ factor is renormalized to $g \approx -0.44$. The ratio of Zeeman energy $g\mu_{B}B$ to the typical Coulomb interaction energy $e^{2}/(\epsilon l_{B})$ is nearly $1/55$ under magnetic field $B=9T$. Thus the Coulomb interaction dominates the Zeeman gap in the magnetic fields of interest. Actually, the $g$ factor could even be tuned by quantum confinement\cite{gfactor} or by application of hydrostatic pressure, and could pass through zero in these circumstances. We thus ignore the Zeeman term when studying the edge spin excitations of integer quantum Hall liquids most of the time. As a result for each energy level with quantum number $S$ obtained in exact diagonalization, it has degeneracy $2 S+1$ with different choices of $S_{z}$. Adding back a Zeeman term will not change the eigenstates, but only split the original degenerate energy levels and give the states with different $S_{z}$ different energy shifts; we will also consider the effects of such splitting below.

\begin{figure}[h]
\includegraphics[width=7cm]{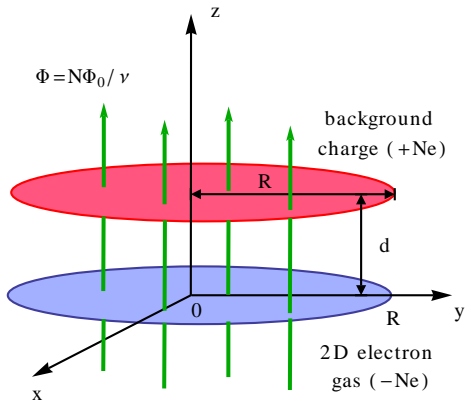}
\caption{(Color online) Sketch of an electron (blue) layer and a uniformly distributed neutralizing background charge (red) layer separated by a distance $d$ from each other in disk geometry.}\label{}
\end{figure}

To model the confinement of 2DEG based on a modulation-doped AlGaAs/GaAs heterostructure, we assume a uniformly distributed neutralizing positive background charge layer at a distance $d$ above the 2DEG as in Fig. 1. The radius of the positive background is $R=\sqrt{2N/\nu}$, so the disk encloses exactly $N/\nu$ magnetic flux quanta for $N$-particle 2DEG system. Coulomb interaction between the positive background charge and the 2DEG gives rise to the {\em positive background charge confining potential}
\begin{equation}
U^{cp}_{m}=\frac{e\rho}{2\pi2^{m}m!}\int\int_{r_{2}<R}d^{2}r_{1}d^{2}r_{2}\frac{1}{\sqrt{d^{2}+r_{12}^{2}}}r_{1}^{2m}e^{-r_{1}^{2}/2},
\label{cpPbg}
\end{equation}
where $\rho$ is the charge density of the background. The ratio of distance $d$ between the electron layer and the background charge layer to magnetic length $l_{B}$: $d/l_{B}$, is the {\em tuning parameter} that controls the relative strength of the confining potential to the electron-electron interaction.

For {\em parabolic confining potential} $U(\mathbf{r})=\beta r^{2} $,
\begin{equation}
\begin{split}
\underset{m,\sigma}{\sum}U_{m}^{cp}\hat{n}_{m,\sigma} &=\underset{m=0,\sigma}{\overset{\infty}{\sum}}\left\langle m\right|\beta r^{2}\left|m\right\rangle \hat{n}_{m,\sigma}  \\
 & =b\underset{m=0,\sigma}{\overset{\infty}{\sum}}2(m+1)(\frac{e^{2}}{\epsilon l_{B}})\hat{n}_{m,\sigma} \\
 & =b\times2(\hat{M}+\hat{N})(\frac{e^{2}}{\epsilon l_{B}}) .
\end{split}
\label{cpParabolic}
\end{equation}
The coefficient $\beta$ represents the steepness of the parabolic confinement, which is in units of $meV/nm^{2}$. In this case,
\begin{equation}
b=\beta l_{B}^{2}/(\frac{e^{2}}{\epsilon l_{B}})
\label{parameterParabolic}
\end{equation}
is the dimensionless {\em tuning parameter}, which is proportional to steepness coefficient $\beta$ and the $3/2$ power of magnetic field strength $B$.
The one body confining potential operator can be written as the function of total orbital angular momentum operator $\hat{M}$, and total particle number operator $\hat{N}$.

For {\em linear confining potential} $U(\mathbf{r})=\alpha r$,
\begin{equation}
\begin{split}U_{m}^{cp}&=\left\langle m\right|\alpha r\left|m\right\rangle  \\
 & =a\sqrt{\frac{\pi}{2}}\frac{(2m+1)!!}{2^{m}m!}(\frac{e^{2}}{\epsilon l_{B}}),
\end{split}
\label{cpLinear}
\end{equation}
where the coefficient $\alpha$ represents the slope of the linear confining potential, which is in units of $meV/nm$. Here,
\begin{equation}
a=\alpha l_{B}/(\frac{e^{2}}{\epsilon l_{B}})
\label{parameterLinear}
\end{equation}
is the dimensionless {\em tuning parameter}, which is proportional to the confining potential's slope $\alpha$ and magnetic field strength $B$.

For the exact diagonalization calculations of small systems below, we limit $N$ particles in a finite number $N_{orb}$ of orbitals, and solve the problem in subspaces with certain quantum numbers $M$, $S$ and $S_{z}$. Therefore the Hilbert space could be limited to a reasonable size and exact diagonalization could be carried out. For the exact diagonalization calculations in Sec. III.A and IV.A, we give each electron system a sufficiently large orbital number $N_{orb}$ ($N_{orb}=16$, $23$, $19$ for $N=8$, $10$, $12$ system respectively), so that the calculated orbital occupation numbers $n_{m} $ approach $0$ in the outermost given orbitals. For this reason, we believe that the limited orbital number $N_{orb}$ we have chosen has little effect to restrict the electrons to the inner orbitals, and the results will not change qualitatively if larger $N_{orb}$ is used in this exact diagonalization calculation.

At this point, we also need to introduce a standard to identify the filling factor $\nu$ of the electron systems for the future calculation. For finite systems it is very easy to identify the $\nu=1$ and $2$ states, because electrons just completely occupy the innermost lowest LL orbitals of the single or two spin component(s) and form a Slater determinant state. The finite size systems for $\nu=2$ state study should have even numbers of particle. As the confinement smoothes, electron system transits from unpolarized $\nu=2$ state to partially polarized states, and finally to polarized $\nu=1$ state. When the system is in this transition region, we say it is in $1\leq\nu\leq2 $ regime. For an electron system with filling factor $\nu=1$, when the confinement becomes smoother (stronger), electrons would tend to distribute in more (less) orbitals. If the confinement becomes a little smoother (stronger) and the system has just undergone a few transitions from the original $\nu=1$ state, we say this system is in $\nu \lesssim 1$ ($\nu \gtrsim 1$) regime. The identification is similar for $\nu=2$ state. But filling factor could not be larger than $2$ if Hilbert space is truncated to the lowest LL, so only $\nu \lesssim 2$ regime exists.

\subsection{Electrostatic consideration of origin of edge instabilities}
\begin{figure}[h]
\includegraphics[width=8.5cm]{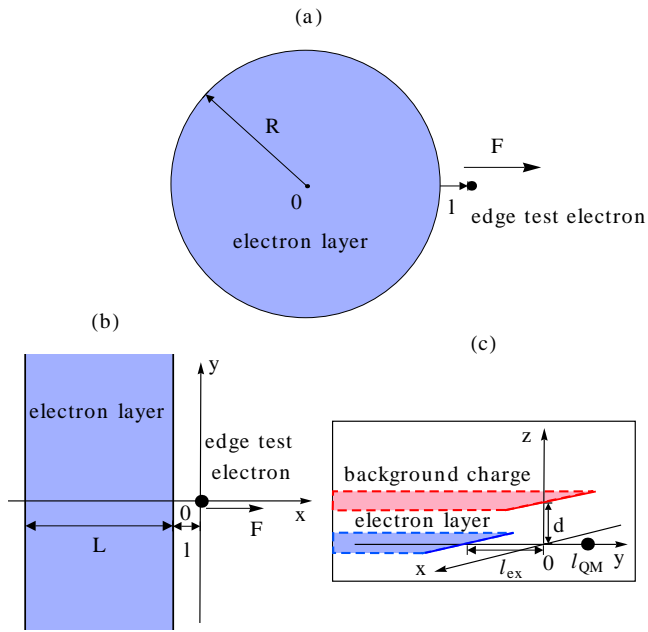}
\caption{(Color online) (a) Edge test electron and electron layer in disk geometry. (b) Edge test electron and eletron layer in strip geometry. (c) Edge test electron and semi-infinite electron layer and positive background charge layer.}\label{}
\end{figure}

Before studying edge instabilities of $\nu = 1$ and $\nu = 2$ QH liquids numerically, we introduce an electrostatic model, which gives us some qualitative understandings of the instabilities, especially for large systems.
The expectation value of $r^{2}$ for single electron state of the lowest LL is $2(m+1)l_{B}^{2}$, so it is reasonable to make the approximation that every $1$ ($2$) electron(s) occupy area $2\pi l_{B}^2$ and the whole $-Ne$ charge uniformly distributes in a disk with radius $R=\sqrt{2N}l_{B}$ ($\sqrt{N}l_{B}$) for $\nu=1$ ($2$) state. The charge density is $\sigma=-e/(2\pi l_{B}^{2})$ for $\nu=1$ state, and $-e/(\pi l_{B}^{2})$ for $\nu=2$ state.

The edge instability and the resultant structure is determined by the competition between electron-electron repulsion and confinement potential. 
We thus consider the forces felt by a test electron at the border of this disk.
Quantum mechanically the electron's wave function has width $\sim l_{B}$ in the radial direction. For this reason, in the electrostatic model, we put the test electron at a point which is $l_{QM}$ away from the electron layer's border and $l_{QM} \sim l_{B}$.
Besides direct interaction, exchange interaction between two electrons with same spin in the 2DEG, is also an essential factor of edge instability. To include exchange interaction in our electrostatic model, we consider the following. First, exchange interaction is attractive, which can be modeled as the direct interaction between some hypothetical {\em positive} charges in the disk and the edge test electron. Second, exchange interaction has a short range with the order of magnetic length $l_{B}$ in the QH regime. Therefore, including the effect of exchange interaction at $\nu=1$ (where exchange effects exists between all pairs of electrons), we reduce the radius of the (negatively charged) disk from $R$ to $R-l_{ex}$ without changing the charge density $\sigma$, where $l_{ex}$ also has the order of $l_{B}$. A similar model (with proper modification) can be used to describe the $\nu=2$ case.

As shown in Fig. 2(a), the electrostatic model for 2DEG $\nu=1$ state is an electron disk with uniform charge density $\sigma=-e/(2\pi l_{B}^{2})$, radius $R=\sqrt{2 N} l_{B}-l_{ex}$, and an edge test electron with distance $l=l_{QM}+l_{ex}$ away from the electron disk. For $\nu=2$ state, we need two electron disks and one edge test electron to model the system; one of the disks with uniform charge density $\sigma=-e/(2 \pi l_{B}^{2})$ and radius $R=\sqrt{N} l_{B}-l_{ex}$, models the electrons in the plane with the same spin as the edge electron, and the other disk with uniform charge density $\sigma=-e/(2 \pi l_{B}^{2})$ and radius $R=\sqrt{N} l_{B}$, models the electrons in the plane with the different spin as the edge electron. Edge test electron is $l=l_{ex}+l_{QM}$ away from the first disk and $l_{QM}$ away from the second disk in the $\nu=2$ state's electrostatic model.

\begin{figure}[h]
\includegraphics[width=5.5cm]{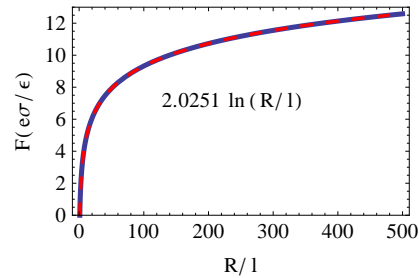}
\caption{(Color online) Numerically calculated force $F$ felt by the test electron {\em vs} $R/l$ in the range $R/l\in[0,500] $ (blue line) and its fitting function $2.0251 \ln(R/l)$ (dashed red line) in disk geometry.}\label{}
\end{figure}

To study the edge instabilities, we first consider the force felt by the edge test electron from the electron disks in this model. In the strip geometry of the same problem, it is easy to obtain an explicit form of this force, which is what we consider first. As Fig. 2(b), it is straightforward to show that the edge test electron feels a force $F=2 \ln(L/l) (e \sigma/\epsilon)$ in $x$ direction from the electron layer, where $L$ is the width of the strip and $l$ is the distance from the test electron to the electron layer. When the system is large ($R\gg l_{B}$), the force felt by edge test electron in disk geometry tends to be the same as the one in strip geometry, where $R$ plays the role of $L$. In disk geometry as Fig. 2(a), force $F$ felt by the edge test electron {\em vs} $R/l$ in range $[0,500]$ is calculated numerically in Fig. 3. It is well fitted by a function $2.0251 \ln(R/l)$. The coefficient of $\ln{R/l}$ approaches $2$ as the strip one with increasing $R$. Therefore, numerically we can see the force felt by edge test electron in disk geometry has a logarithmic dependence of $R/l$. $l$ equals $l_{QM}+l_{ex}$ as the test electron's spin is the same as the ones in electron disk, and $l_{QM}$ as they are different.

The inward force from the confining potential is proportional to $R$ under parabolic confining potential, and is a constant under linear confining potential. Because of logarithmical $R$ dependence of the outward force from the electron disk, thermodynamic limit does not exist for edge instability critical points ($b_c$ and $a_c$) for these two types of confinements.
By further approximating that $l_{QM}\approx l_{B}$, $l_{ex}\approx 0$ and $F\approx 2 \ln(R/l)$, the electrostatic model predicts that for $N\gg 1$ the critical point of $\nu=1$ instability (smoother confinement side) decreases (increases) as $b_c \sim [\ln(2N)]/(4 \pi \sqrt{2N})$ ($a_c\sim [\ln(2N)]/(2 \pi)$) with increasing particle number $N$ under parabolic (linear) confining potential. By the same approximations for $N\gg 0$, the critical point of $\nu=2$ instability decreases (increases) as $b_c \sim [\ln N]/(2 \pi \sqrt{N})$ ($a_c \sim (\ln N)/\pi$) with increasing particle number $N$ under parabolic (linear) confining potential.

For large systems under positive background charge confining potential, the electron layer and background charge layer could be approximated as two semi-infinite layers as Fig. 2(c). For $\nu=1$ state, when the distance of the two layers $d=\sqrt{l_{ex}^{2}+2 l_{ex} l_{QM}}$, the force felt by the edge test electron becomes zero and $\nu=1$ edge instability happens when $d$ further increases. For $\nu=2$ state, when the distance of the two layers $d=\sqrt{l_{ex}^{2}/4+l_{ex} l_{QM}}$, the force felt by the edge test electron becomes zero and $\nu=2$ edge instability happens when $d$ further increases. Because $l_{QM}$ and $l_{ex}$ are independent of $R$ when $R \gg l_B$, the thermodynamic limit exists for positive background charge confinement. From the electrostatic model, we can also predict that in thermodynamic limit, the $\nu=2$ edge instability critical point $d_{c}$ is larger than the one for $\nu=1$ edge instability, and they are both of order $l_{B}$.

These conclusions are confirmed by numerical results presented below, and also consistent with earlier work on fractional QH edge reconstructions.\cite{Wan}

\section{LOW ENERGY EDGE EXCITATIONS AND EDGE INSTABILITIES OF $\nu=1$ FERROMAGNETIC STATE}
In this section, we investigate the low energy excitations of $\nu=1$ ferromagnetic state under the three kinds of confinements. First exact diagonalization results are presented for small systems. The low energy spectra can be mapped onto those of $\Delta S=-1$ bosons on top of the ferromagnetic states. By investigating the spin structure of these low energy excitations, we conclude that they are edge spin waves (ESWs). Microscopic trial wave functions are constructed to describe the ESWs, which allow for their studies in larger systems. We conclude that the edge instabilities of the $\nu=1$ ferromagnetic state are triggered by the condensation of the ESWs which results in edge spin textures.

\subsection{Exact diagonalization study for small systems}
\begin{figure}[h]
\includegraphics[width=8.5cm]{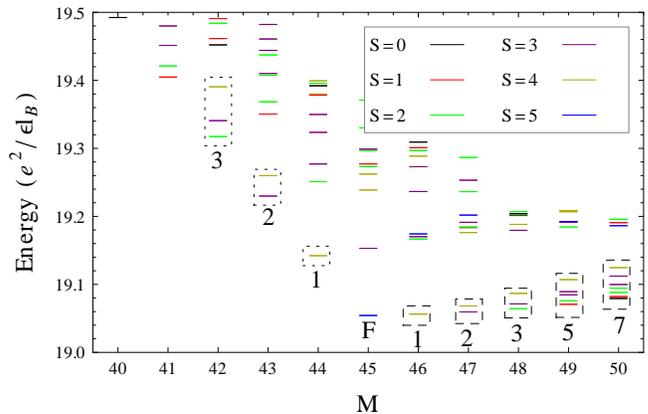}
\caption{(Color online) Low energy spectrum of $10$-electron system under parabolic confining potential when $b=0.06$. The different total spin quantum numbers $S$ of the eigenstates are labeled by different colors as the annotation. The ground state is $\nu=1$ ferromagnetic state (F). Low energy excitations in $\Delta M>0$ subspaces are enclosed by dashed boxes, and low energy excitations in $\Delta M<0$ subspaces are enclosed by dotted boxes. The number below each box is the number of states inside the box. }\label{}
\end{figure}

\begin{table}[h]
\begin{tabular}{|c|c|c||c|c|c|}
\hline \hline
$\Delta M$ & boson configuration & $\Delta S$ & $\Delta M$ & boson configuration & $\Delta S$\\ \hline
$\pm 1$ & $n_{\pm1}=1$ & 1 & \multirow{2}{*}{$\pm 4$} & $n_{\pm2}=2$ & 2 \\
\cline{1-3} \cline{5-6}
\multirow{2}{*}{$\pm 2$} & $n_{\pm1}=2$ & 2 &  & $n_{\pm4}=1$ & 1 \\
\cline{2-6}
 & $n_{\pm2}=1$ & 1 & \multirow{5}{*}{$\pm 5$} & $n_{\pm1}=5$ & 5 \\
\cline{1-3} \cline{5-6}
\multirow{3}{*}{$\pm 3$} & $n_{\pm1}=3$ & 3 &  & $n_{\pm1}=3$, $n_{\pm2}=1$ & 4 \\
\cline{2-3} \cline{5-6}
 & $n_{\pm1}=1$, $n_{\pm2}=1$ & 2 &  & $n_{\pm1}=2$, $n_{\pm3}=1$ & 3 \\
\cline{2-3} \cline{5-6}
 & $n_{\pm3}=1$ & 1 &  & $n_{\pm1}=1$, $n_{\pm2}=2$ & 3 \\
\cline{1-3} \cline{5-6}
\multirow{3}{*}{$\pm 4$} & $n_{\pm1}=4$ & 4 &  & $n_{\pm1}=1$, $n_{\pm4}=1$ & 2 \\
\cline{2-3} \cline{5-6}
 & $n_{\pm1}=2$, $n_{\pm2}=1$ & 3 &  & $n_{\pm2}=1$, $n_{\pm3}=1$ & 2 \\
\cline{2-3} \cline{5-6}
 & $n_{\pm1}=1$, $n_{\pm3}=1$ & 2 &  & $n_{\pm5}=1$ & 1 \\
\hline \hline
\end{tabular}
\caption{One-to-one correspondence between the configurations of spin $\Delta S=-1$ bosons and the low energy excitations in each angular momentum subspace $\Delta M$ of $N=10$ system under parabolic confinement. $\Delta M$ is the difference of angular momentum quantum number $M$ compared to the one for ferromagnetic state. $n_{q}$ is the number of bosons with angular momentum $q$.}
\label{tab:myfirsttable}
\end{table}

\begin{figure}[h]
\includegraphics[width=7cm]{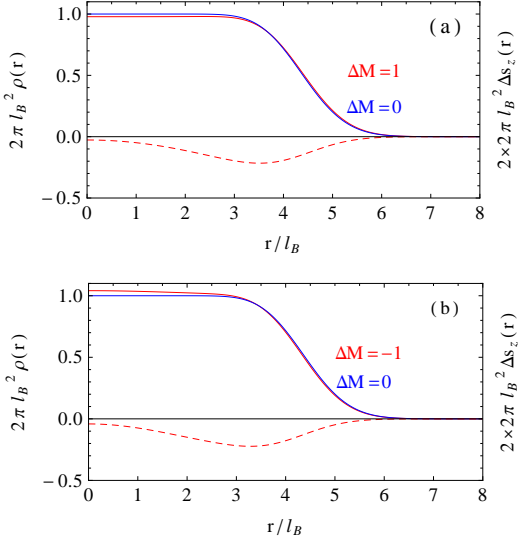}
\caption{(Color online) Charge density $\rho(r)$ (red lines) and change of spin density's z axis component $\Delta s_{z}(r)$ (dashed red lines) for the low energy excitations of $\nu=1$ ferromagnetic state, in (a) subspace $\Delta M=1$, $S_{z}=S=4$ and (b) subspace $\Delta M=-1$, $S_{z}=S=4$, of the $10$-electron system under parabolic confinement with $b=0.06$.
The charge density profiles of low energy excitations have no big change compared to the one of ferromagnetic state (blue line).
}\label{}
\end{figure}

\begin{figure}[h]
\includegraphics[width=8.5cm]{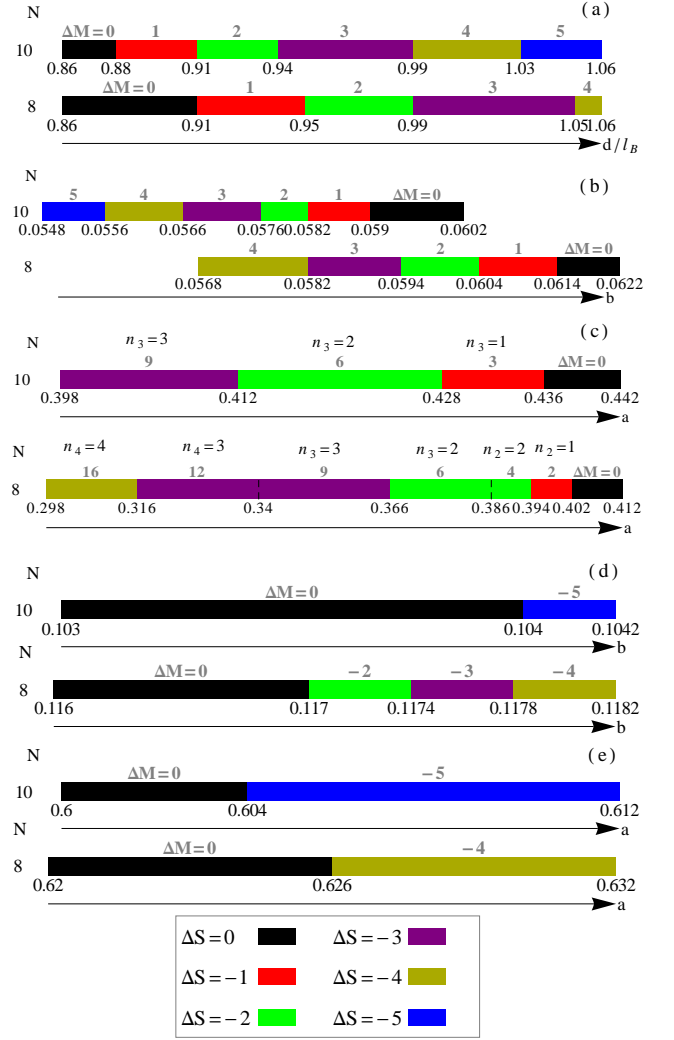}
\caption{(Color online) Phase diagrams of $8$, $10$-electron systems under (a) $\nu=1$ positive background charge, (b) parabolic, (c) linear confining potential in $\nu\lesssim 1$ regime, and under (d) parabolic, (e) linear confining potential in $\nu\gtrsim 1$ regime based on exact diagonalization. $\Delta M$ ($\Delta S$) is the difference between orbital angular momentum (spin) quantum number $M$ ($S$) and that of $\nu=1$ ferromagnetic state for each system. $\Delta S$ of ground states are labeled by the bars with different colors as the annotation. Step values are $\Delta (d/l_{B})=10^{-2}$, $\Delta b=2\times 10^{-4}$ and $\Delta a=2\times 10^{-3}$ for the three confinements respectively. The configuration of bosons of case (c) is written above each ground state and $n_{q}$ is the occupation number of the boson with angular momentum $q$. For the ground states of all the other cases above, the configurations of bosons are $n_{1}=\Delta M$ if $\Delta M>0$, and $n_{-1}=-\Delta M$ if $\Delta M<0$. }\label{}
\end{figure}

In this subsection exact diagonalization method is used to study the small electron systems close to $\nu=1$ ferromagnetic state under the three confinements. Fig. 4 is the $10$-electron system's low energy spectrum under parabolic confinement with $b=0.21$ (ferromagnetic state is the ground state). We will use this specific spectrum to show some low energy spectra's common characteristics shared by all small systems we studied and under all the three confinements. $\Delta M=M-M_{0}$ is the difference of orbital angular momentum quantum number $M$ compared to the one $M=M_{0}$ for ferromagnetic state ($M_{0}=45$ for $N=10$ system). In the subspaces with $\Delta M>0$, a set of low energy excitations (enclosed by dashed boxes in Fig. 4) are separated from higher energy states. One of these excitations becomes the ground state and thus destabilizes the ferromagnetic ground state with smoother confinement. These low energy excitations all have nonzero $\Delta S$ and the polarized excitations with $\Delta S=0$ have higher energies compared to them. In the subspaces with $\Delta M<0$, another set of low energy excitations (enclosed by dotted boxes in Fig. 4) are separated from higher energy states, and one of these excitations becomes the ground state with stronger confinement (except for positive background charge confinement case where the ferromagnetic state remains stable all the way to $d=0$.). Polarized excitation with $\Delta S=0$ does not exist in the subspaces with $\Delta M<0$, because all the lowest orbitals with up spin are already occupied in the ferromagnetic state.
More interestingly, the low energy spectrum (enclosed by the boxes) matches that of a bosonic system, with each boson reduces the system's total spin by $1$. The one-to-one correspondence between the bosonic states and low energy excitations of this $10$-electron system under parabolic confinement is shown in Table I.
Figs. 5 shows the charge and spin density profiles of the bosonic states in subspaces $\Delta M=1$ and $-1$ of the spectrum Fig. 4. While charge density profiles are almost identical to that of the ferromagnetic state, we find the flipped spin is confined to the edge. We thus conclude the bosons excited from the ferromagnetic state are actually edge spin waves (ESWs),\cite{Sondhi99} which can propagate both along the clockwise ($\Delta M>0$ bosons) and anti-clockwise ($\Delta M<0$ bosons) directions.
We repeat that the characteristics stated above are shared by all small systems we studied and under all the three confinements.

Fig. 6 presents the phase diagrams of $8$, $10$-electron systems close to $\nu=1$ ferromagnetic states under the three confinements. As the confinement becomes smoother, the energy of a single boson state with some positive angular momentum $q=q_{0}$ corresponding to clockwise ESW first crosses that of the ferromagnetic state, thus destabilizes it. There are interactions among these bosons. As the confinement further smoothes, the bosons start to condense because in this process the benefit of adding a new boson with angular momentum $q_{0}$ would exceed the cost of the accompanying increase of interaction energy. When the confinement becomes smoother, the minimum of boson dispersion shift from $q=q_{0}$ to $q=q_{0}'$. If that happens, the system will choose to condense $q_{0}'$ bosons, instead of $q_{0}$ bosons. Using this picture, we can understand why in the phase diagrams Fig. 6(a-c) the systems flip spins one by one with smoother confinement and quantum number $\Delta M+q_{0} \Delta S$ is conserved for all the ground states. $q_{0}$ may change along with different tuning parameters as in Fig. 6(c).
When the confinement becomes stronger, the systems start to condense bosons with $q=-1$, which correspond to anti-clockwise ESWs. As in Figs. 6(d,e), the systems will condense all the possible $q=-1$ bosons to form spin singlet states either at one time, or within a much smaller range of tuning parameter compared to the transitions in smoother confinement side.

Exact diagonalization results also show that under positive background charge confining potential, once the system transits to $\nu=1$ ferromagnetic state, the phase will be stable as $d/l_{B}$ approaches zero.

When the confinement becomes smoother, for all the three confinements $\delta U_{m}^{cp}<\delta U_{m'}^{cp}$ ($m>m'$), where $\delta U_{m}^{cp}$ is the change of confining potential in orbital $m$. The electrons near the edge see an larger additional outward force, thus are more likely to jump to the ($m+q$)th orbital ($q>0$) with down spin from its original $m$th orbital with up spin to form a particle-hole pair $b_{m,q,\downarrow}^{\dagger}\left|F\right\rangle $. By contrast, when the confinement becomes stronger, $\delta U_{m}^{cp}>\delta U_{m'}^{cp}$ ($m>m'$). The electrons in the bulk see an larger additional inward force, thus are more likely to jump to the ($m+q$)th orbital ($q<0$) with down spin from its original $m$th orbital with up spin to form a particle-hole pair. Therefore, in $\Delta S_{z}=-1$ sector, all bosonic excitations with positive $q$ happen near the edge (ESW); while all bosonic excitations with negative $q$ happen in the bulk (SW).

\subsection{Microscopic trial wave function study for large systems}
Having identified the low energy excitations at $\nu=1$ as bosonic ESWs on top of the ferromagnetic state, we want to construct microscopic trial wave functions to approximate them. Like Abelian bosonization in one-dimensional Fermi gas, the bosonic generators do not respect the SU($2$) spin rotational symmetry. For this reason, we only choose one state among all the degenerate states in each energy level with spin quantum number $S$ to represent all states in this level. The state we will choose is the one with $S_{z}=S$, because upon adding a finite Zeeman term this state's energy is the lowest within this family.

On top of the $S_{z}=S$ ferromagnetic state
\begin{equation}
\left|F\right\rangle =\overset{N-1}{\underset{m=0}{\prod}}c_{m,\uparrow}^{\dagger}\left|0\right\rangle,
\label{ferroState}
\end{equation}
the bosonic generators of ESWs were first constructed by Oaknin et al.\cite{Oaknin} We rewrite them as
\begin{equation}
B_{q,\sigma}^{\dagger} =\overset{N-1}{\underset{m=0}{\sum}}\psi_{m,q,\sigma}b_{m,q,\sigma}^{\dagger}  ,
\label{trialWf}
\end{equation}
where $b_{m,q,\sigma}^{\dagger}\equiv c_{m+q,\sigma}^{\dagger}c_{m,\uparrow}$, $q$ is an integer and $\sigma=\uparrow$ or $\downarrow$. There is one flipped spin if $\sigma=\downarrow$ ($\Delta S_{z}=-1$), and no flipped spin if $\sigma=\uparrow$ ($\Delta S_{z}=0$). On top of the ferromagnetic state with $S_{z}=S$ in Eq. (\ref{ferroState}), $b^{\dagger}_{m,q,\sigma}$ generate a basis with dimension $N$ for $\sigma=\downarrow$ and dimension $q$ for $\sigma=\uparrow$. After diagonalizing the matrix 
\begin{equation}
\left\langle F\right|\left(\begin{array}{c}
b_{0,q,\sigma}\\
b_{1,q,\sigma}\\
\vdots\\
b_{N-1,q,\sigma}
\end{array}\right)H\left(\begin{array}{cccc}
b_{0,q,\sigma}^{\dagger}, & b_{1,q,\sigma}^{\dagger}, & \cdots, & b_{N-1,q,\sigma}^{\dagger}\end{array}\right)\left|F\right\rangle  
\label{trialWfMatrix}
\end{equation}
in which the Hamiltonian $H$ is defined in Eq. (\ref{completeH}) and $g=0$, $\psi_{m,q,\sigma}$ is determined by identifying that $\overset{N-1}{\underset{m=0}{\sum}}\psi_{m,q,\sigma}b_{m,q,\sigma}^{\dagger}$ is the state with the lowest energy and is normalized. Because
\small
\begin{equation}
[B_{q,\sigma},B_{q',\sigma'}^{\dagger}]=\delta_{q,q'}\delta_{\sigma,\sigma'}\underset{m=0}{\overset{N-1}{\sum}}\psi_{m,q,\sigma}^{2}(b_{m,q,\sigma}b_{m,q,\sigma}^{\dagger}-b_{m,q,\sigma}^{\dagger}b_{m,q,\sigma})
\label{Bcommute}
\end{equation}
\normalsize
but
\begin{equation}
\left\langle F\right| [B_{q,\sigma},B_{q',\sigma'}^{\dagger}]\left|F\right\rangle=\delta_{q,q'}\delta_{\sigma,\sigma'} ,
\label{BosonCommute}
\end{equation}
the states generated by $B_{q,\sigma}^{\dagger}$ can be approximated as bosons, especially when the number of excitations are small.
The state of $n$ bosons with $q$ and $\sigma$ is
\begin{equation} 
(B_{q,\sigma}^{\dagger})^{n}\left|F\right\rangle =(\overset{N-1}{\underset{m=0}{\sum}}\psi_{m,q,\sigma}c_{m+q,\sigma}^{\dagger}c_{m,\uparrow})^{n}\left|F\right\rangle .
\label{nBoson}
\end{equation}

\begin{table}[h]
\begin{tabular}{|c|c|c|c|}
\hline \hline
$N$ & positive background &\qquad parabolic \qquad \quad &\qquad linear \qquad \quad \\ \hline
8 & 0.8643 & 0.8584 & 0.8865\\ \hline
9 & 0.9164 & 0.9136 & 0.9134\\ \hline
10 & 0.9889 & 0.9896 & 0.9530\\ \hline
11 & 0.9904 & 0.9910 & 0.9864\\ \hline
12 & 0.9916 & 0.9920 & 0.9875\\ \hline
13 & 0.9926 & 0.9929 & 0.9884 \\ \hline
14 & 0.9934 & 0.9936 & 0.9891\\ \hline
15 & 0.9941 & 0.9942 & 0.9897\\ \hline
16 & 0.9946 & 0.9947 & 0.9901\\ \hline
17 & 0.9951 & 0.9951 & 0.9905\\ \hline
18 & 0.9955 & 0.9954 & 0.9909\\ \hline
19 & 0.9959 & 0.9957 & 0.9912\\ \hline
20 & 0.9962 & 0.9960 & 0.9914\\ \hline
\hline
\end{tabular}
\caption{Overlaps between $B_{1,\downarrow}^{\dagger}\left|F\right\rangle  $ and the exact lowest energy state in the subspace with $M=M_{0}+1$ and $S_{z}=N/2-1$, as functions of the number of electrons under three different kinds of confinements. In calculating these states, tuning parameters of confinement are chosen as the critical values of the ferromagnetic state's initial instibility with smoother confinement.}
\label{tab:myfirsttable}
\end{table}

Compared with the exact low energy excitations, the microscopic trial wave functions (\ref{nBoson}) have good quantum numbers $M$ and $S_{z}$ but not $S$. Nevertheless they do have large overlaps with the exact excited states (see Table II for examples). The overlaps {\em increase} with increasing particle number $N$, and we do not fully understand the reason of this surprising behavior yet.

The bosonic excitations (\ref{nBoson}) are classified into two different types. The excitations generated by $B_{q,\uparrow}^{\dagger}$ ($\Delta S_{z}=0$ sector) are the well studied edge magneto-plasmons (EMPs).\cite{EricYang1, Chamon} Since $B_{q,\uparrow}^{\dagger}|F\rangle=0$ for $q < 0$, EMP can only carry {\em positive} angular momentum and is thus chiral. It excites the edge by creating a charge density wave with positive angular momentum $q$ which propagates along the clockwise direction {\em only}. The excitations generated by $B_{q,\downarrow}^{\dagger}$ ($\Delta S_{z}=-1$ sector) are the ESWs discussed in last subsection. As a non-chiral excitation, ESW excites the edge by creating a clockwise edge spin wave with positive $q$ or anti-clockwise edge spin wave with negative $q$.

For larger systems, exact diagonalization is not feasible. The microscopic wave functions (\ref{nBoson}) give us a tool to study the instabilities near the ferromagnetic states.
The energies of a single boson with $q$, $\sigma$ and two bosons with the same $q$, $\sigma$ on top of the ferromagnetic state are
\begin{equation}
\triangle E_{1}^{(q,\sigma)}=\frac{\left\langle F\right|B_{q,\sigma}HB_{q,\sigma}^{\dagger}\left|F\right\rangle}{\left\langle F\right|B_{q,\sigma}B_{q,\sigma}^{\dagger}\left|F\right\rangle} -\left\langle F\right|H\left|F\right\rangle
\label{firstE}
\end{equation}
and
\begin{equation}\triangle E_{2}^{(q,\sigma)}=\frac{\left\langle F\right|(B_{q,\sigma})^{2}H(B_{q,\sigma}^{\dagger})^{2}\left|F\right\rangle}{\left\langle F\right|(B_{q,\sigma})^{2}(B_{q,\sigma}^{\dagger})^{2}\left|F\right\rangle} -\left\langle F\right|H\left|F\right\rangle
\label{secondE}
\end{equation}
respectively. $\varepsilon^{(q,\sigma)}_{0}=\triangle E_{1}^{(q,\sigma)}$ is the kinetic energy of a single boson, and $\varepsilon_{int}^{(q,\sigma)}=\Delta E_{2}^{(q,\sigma)}-2 \Delta E_{1}^{(q,\sigma)}$ is the interaction energy between the two bosons. The energy of $n$ bosons with the same $q$ and $\sigma$ on top of the ferromagnetic state is
\begin{equation}
\bigtriangleup E_{n}^{(q,\sigma)}=n\varepsilon_{0}^{(q,\sigma)}+\frac{n(n-1)}{2}\varepsilon_{int}^{(q,\sigma)}.
\label{nthE}
\end{equation}

In the exact diagonalization calculation for small systems, besides ferromagnetic state, the ground states are always the bosonic states consisting of bosons with the same $q$ and $\sigma$.
Assuming this is also true for larger systems and using Eq. (\ref{nthE}), we obtain the phase diagrams close to ferromagnetic states under the three confinements for particle number up to $40$ (Fig. 7). We first notice that when the confinement smoothes, phase boundaries of ferromagnetic states are determined by the ESW instabilities, rather than the EMP instabilities. With smoother confinement, ESWs are generated one by one to lower the systems' energy.

\begin{figure}[h]
\includegraphics[width=5.7cm]{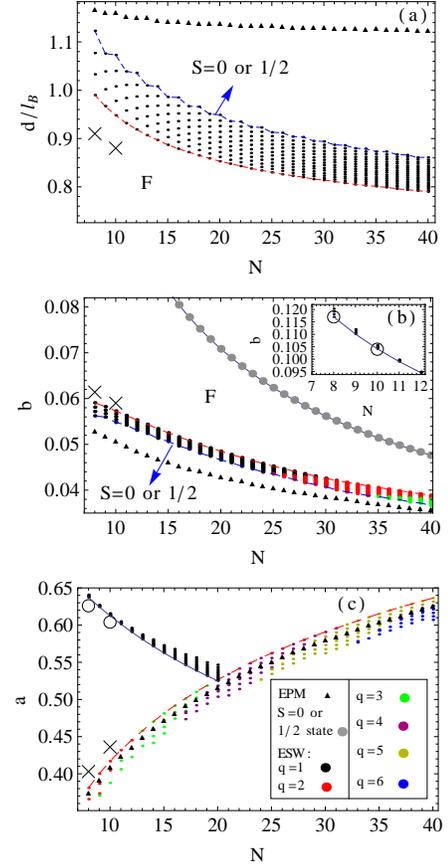}
\caption{(Color online) Phase diagrams based on trial wave functions close to $\nu=1$ for large systems under (a) $\nu=1$ positive background charge, (b) parabolic and (c) linear confining potentials. $\nu=1$ ferromagnetic states (F) transit to bosonic states of the form $(B_{q,\downarrow}^{\dagger})^{n}\left|F\right\rangle $ with varying strength of confinement. Whenever one more boson is generated, a point is plotted to label the critical tuning parameter of this transition. The color of each point represents the angular momentum quantum number $q$ of generated bosons, as labeled in the annotation. There is one exception. Under parabolic confinement and with stronger confinement, the ferromagnetic state condenses $q=-1$ bosons one by one when $N\leq 12$, but transits to unpolarized (even $N$) or spin $1/2$ (odd $N$) state directly when $N>12$ (its critical point is labeled by a bigger gray point). In the smoother confinement side, the phase boundaries of ferromagnetic states are determined by the edge spin wave instabilities, rather than the edge magnetoplasmon instabilities (labeled by triangles). For $8$ and $10$-electron systems, the critical points of ferromagnetic state's initial instability calculated by exact diagonalization are labeled by crosses in smoother confinement side, and circles in stronger confinement side.
}\label{}
\end{figure}

\begin{figure}[h]
\includegraphics[width=6cm]{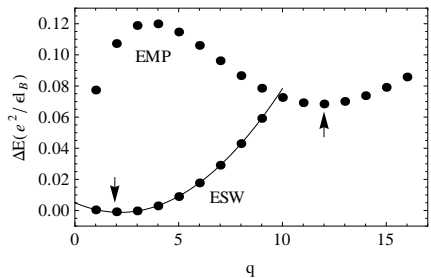}
\caption{Dispersion relation calculated by trial wave functions for edge magnetoplasmon (EMP) and edge spin wave (ESW) in $N=40$ system under parabolic confining potential ($b=0.0387$). The arrows label the minimum of each branch of excitations. The dispersion relation of ESW is fitted well by a parabola.
}\label{}
\end{figure}

For positive background charge confining potential, ESWs in $\Delta S_{z}=-1$ sector with $q=1$ are generated on the ferromagnetic state one by one with smoother confinement. The critial value of ferromagnetic state's initial instability approaches a constant with increasing particle number, which implies the existence of thermodynamic limit. Therefore, we predict that in thermodynamic limit, the instabilities close to ferromagnetic state under positive background charge confinement are the condensation of $\Delta S_{z}=-1$ bosons with modulation angular momentum $q=1$ (the lowest possible modulation angular momentum), and the critial point of ferromagnetic state's initial instability $d_{c}/l_{B}\approx 0.8$.

Under parabolic and linear confinements, the systems tend to condense $\Delta S_{z}=-1$ bosons with larger modulation angular momentum $q$ to minimize the energy, especially when the particle number becomes large. For a certain system, the minimum of the bonson's dispersion may also change from one angular momentum $q$ to another. Fig. 8 shows the dispersion relation of ESW in $N=40$ system under parabolic confinement when $b=0.0387$. At this tuning parameter, the minimum is at $q=2$ and its energy is lower than the ferromagnetic state's energy, so the system prefers to condense $q=2$ bosons. With smaller $b$, the minimum would shift to $q=3$, then the system gives up the $q=2$ bosons and chooses to condense $q=3$ bosons. This process can also be observed in the phase diagram Fig. 7(b), where the color of instabilities' critical points changes from red to green with smaller $b$ when $N=40$.
As the confinement is strengthened under parabolic and linear confinement, $\Delta S_{z}=-1$ bosons with $q=-1$ are always favored to be condensed. It is the condensation of these ESWs that drives spin edge reconstruction (or formation of edge spin textures) at $\nu=1$.\cite{note} First order phase transition will happen under parabolic confinement ($N>12$). Instead of generating bosons one by one as the linear confinement case or small systems ($N \leq 12$), the system prefers to generate all the possible bosons at one time. The window of $\nu=1$ ferromagnetic state decreases to zero with increasing particle numbers under parabolic and linear confinement. Thus the $\nu=1$ ferromagnetic state will not appear as ground state in large systems under these two confinements. Also consistent with the electrostatic analysis, the critical points of edge instabilities in the smoother confinements side decrease (increase) with increasing particle number under parabolic (linear) confining potential. For $8$ and $10$-electron systems, our trial wave function method's results are in quantitative agreement with exact diagonalization results of Sec. III.A; the latter are labeled in Fig. 7 for comparison.

\begin{figure}[h]
\includegraphics[width=8.5cm]{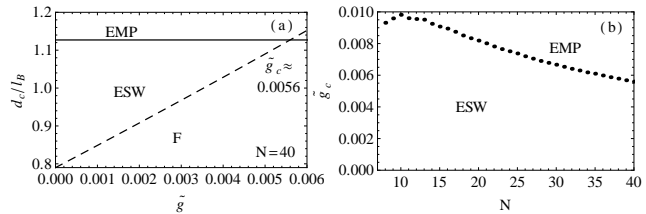}
\caption{(a) Stability diagram calculated by trial wave function for the $40$-electron system's ferromagnetic state (F) under positive background charge confining potential, showing the region of stability to the ferromagnetic state, and the regions where it is unstable to edge spin wave (ESW) and edge magneto-plasmon (EMP) instabilities.
(b) Renormalized critical $g$ factor $\tilde{g_{c}}=g_{c} \mu_{B} B/(e^{2}/\epsilon l_{B})$ for the ferromagnetic state having EMP, instead of ESW initial instability {\em vs} particle number $N$ under positive background charge confining potential calculated by trial wave function.
}\label{}
\end{figure}
Finite Zeeman term will lift the energies of ESWs. The dimensionless parameter $\tilde{g}=g \mu_{B} B/(e^{2}/\epsilon l_{B}) $ is the ratio of the Zeeman energy to the typical Coulomb energy. For large enough $\tilde{g}=\tilde{g_{c}}$ the ferromagnetic state will prefer to transit to its polarized excitations, and EMP will replace ESW to become the initial instability of ferromagnetic state. Fig. 9(a) shows the stability diagram of the $40$-electron system under positive background charge confinement in ($\tilde{g}$, $d/l_{B}$) plane with phase boundaries obtained by our trial wave function. For small Zeeman energies, $\tilde{g}<\tilde{g_{c}}=0.0056$, the ferromagnetic state is destabilized by the ESWs with smoother confinement. As $\tilde{g_{c}}$ increases, the critical tuning parameter $d_{c}/l_{B}$ of ferromagnetic state's initial instability increases as well until at $\tilde{g_{c}}=0.0056$, the polarized EMP becomes the initial instability. Fig. 9(b) shows how the critical value $\tilde{g_{c}}$ changes with increasing particle number. With thermodynamic limit under positive background charge confinement $\tilde{g_{c}}$ approaches a constant with increasing particle number. Based on this figure, we predict that in thermodynamic limit, $\tilde{g_{c}}\approx0.005$. And if we choose dielectric constant $\epsilon=12.8$, $g=-0.44$ for the 2DEG based on a modulation-doped AlGaAs/GaAs heterostructure, the critical magnetic field magnitude is $B_{c}=0.74$T corresponding to the renormalized critical $g$ factor $\tilde{g_{c}}=0.005$. Our quantitative prediction of $\tilde{g_{c}}$ is about one half that obtained by J. Sjostrand {\em et al.},\cite{Sjostrand} who used a modified version of the positive background charge confinement which is slightly sharper. Other results like $d_c$ also differ by a order 1 numerical factors. These indicate such quantitative results are sensitive to details of the confining potential. The fact that $\tilde{g_{c}}$ is larger for sharper confinement is not surprising as sharper confinement tends to lift the energy of charge modes more than spin modes. We note that in real samples there are usually additional sources of confining potential on top of the background charge, like gates and in particular, sharp crystalline boundaries. Thus in reality $B_{c}$ may well exceed $\sim 1$T. Also in reality $d\ll d_c$. As a result edge spin textures (possibly in combination with charge reconstruction) are likely to be present in many systems.

\section{LOW ENERGY EDGE EXCITATIONS AND EDGE INSTABILITIES OF $\nu=2$ UNPOLARIZED STATE}
In this section, we investigate the low energy excitations and edge instabilities of $\nu=2$ unpolarized state under the three kinds of confinements. Carrying out exact diagonalization on small systems in $1\leq\upsilon\leq2 $ regime, we observe that compact states (defined below) with total spin from $0$ to $N/2$ become ground states in some regions of tuning parameter. Similar to the ferromagnetic state case, low energy spectra can be mapped onto those of the bosons on top of the $\nu=2$ unpolarized state. The non-compact ground states with simultaneous edge spin and charge reconstruction are understood as the condensation of bosons (ESWs) on top of $S\neq 0$ compact states.
Larger systems are studied by separating edge electrons from bulk ones; this allows us to reach conclusions about thermodynamic limit in certain cases.

\subsection{Exact diagonalization study of small systems}
\begin{figure*}[t]
\includegraphics[width=13cm]{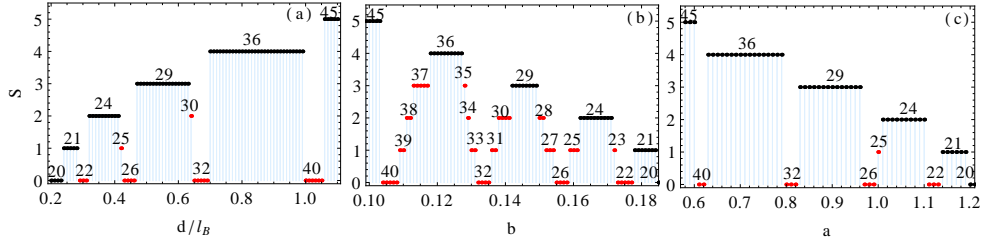}
\caption{(Color online) Phase diagrams for $10$-electron systems under (a) $\nu=2$ positive background charge, (b) parabolic, (c) linear confining potential in $1\leq \nu \leq 2$ regime. Step values are $\Delta (d/l_{B})=10^{-2}$, $\Delta b=10^{-3}$ and $\Delta a=10^{-2}$ for the three confinements respectively. Compact states are labeled by black points, while non-compact states are labeled by red points. The total orbital angular momentum number $M$ of each state is labeled above the points. See text for the definition of compact state.}\label{}
\end{figure*}
\begin{figure}[h]
\includegraphics[width=4.5cm]{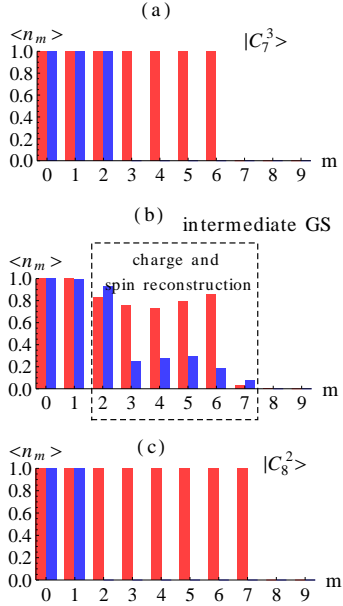}
\caption{(Color online) Occupation numbers $\left\langle n_{m}\right\rangle  $ of (a) compact state $\left|C_{7}^{3}\right\rangle$, (b) non-compact ground state (GS) with $M=25$, $S=1$ appearing between $\left|C_{7}^{3}\right\rangle$ and $\left|C_{8}^{2}\right\rangle$ in Fig. 10(b) and (c) compact state $\left|C_{8}^{2}\right\rangle$. Red and blue bars label the occupation numbers of spin up and down orbitals respectively.}\label{}
\end{figure}
In this subsection, we study small systems in $1\leq \nu \leq 2$ regime. Exact diagonalization is carried out in $8$, $10$, $12$-electron systems, and the phase diagrams of $10$-electron system (Fig. 10) are used to show some common characteristics shared by all small systems we studied. Starting from $\nu=2$ unpolarized state (with $M=20$ in Figs. 10), with smoother confinement, the compact states with spin $S$ from $0$ to $N/2$ become ground states (labeled by black points) in some tuning parameter regions. A compact state is the state having the minimum $M$ compatible with a given $S$ (the degeneracy is $2S+1$ without Zeeman coupling). The compact state with $S_{z}=S$ is
\begin{equation}
\left|C_{N-k}^{k}\right\rangle =\overset{N-k-1}{\underset{m=0}{\prod}}c_{m,\uparrow}^{\dagger}\overset{k-1}{\underset{m'=0}{\prod}}c_{m',\downarrow}^{\dagger}\left|0\right\rangle  .
\label{compactWf}
\end{equation}
It is a single Slater determinant with the $N-k$ lowest $m$ single particle states with spin up and the $k$ lowest $m$ single particle states with spin down occupied, in which $k$ is an integer running from $0$ to $N/2$. They are the ground states obtained in previous work based on HF approximation.\cite{Dempsey, Yacoby}
The fact that such compact states occupy large portions of the phase diagrams in our exact diagonalization study provides strong support to earlier HF studies.
The compact states with $S_{z}<S$ can be obtained using spin lowering operator
\begin{equation}
S^{-}=\underset{m=1}{\overset{\infty}{\sum}}c_{m,\downarrow}^{\dagger}c_{m,\uparrow}
\label{SLower}
\end{equation}
on the $S_{z}=S$ compact state $\left|C_{N-k}^{k}\right\rangle$.
Besides compact states, other ground states with smaller $S$ (labeled by red points in Figs. 10) appear between neighboring compact states. These non-compact ground states have charge and spin textures, and will be studied in the next subsection. We find that the difference between a non-compact ground state and the neighboring compact states $\left|C_{N-k}^{k}\right\rangle$ and $\left|C_{N-k+1}^{k-1}\right\rangle$ lies mostly  in the occupation numbers between the ($k-1$)th and the ($N-k$)th orbital (see Fig. 11 as an example); these orbitals are occupied by spin up electrons only in state $\left|C_{N-k+1}^{k-1}\right\rangle$. The non-compact ground states resemble HF states with simultaneous charge and spin reconstructions.\cite{Yafis}

\subsection{Separation between bulk and edge electrons and exact diagonalization study of large systems}
Through our exact diagonalization study for small systems in $\nu\leq 2$ regime, we realize that with smoother confinement the compact states with increasing total spin become ground states in some tuning parameter regions; the ground states appear in between compact state $\left|C_{N-k}^{k}\right\rangle$ and $\left|C_{N-k+1}^{k-1}\right\rangle$ have their major charge and spin reconstructions in the same orbital segment occupied by single spin electrons in compact state $\left|C_{N-k+1}^{k-1}\right\rangle$. Assuming these two points still hold for larger systems, when studying the low energy edge excitations and edge instabilities of $\nu=2$ unpolarized state of large system, we can use the exact diagonalization to study a separated edge electron system because the occupation number of the orbitals inside this edge system (bulk orbitals) are nearly $2$ (completely occupied). The effective Hamiltonian for the edge electron system in $\nu\lesssim 2$ regime is
\begin{equation}
\begin{split}
H_{eff}=&\frac{1}{2}\underset{m\geq m_{0},n\geq m_{0},l,\sigma,\sigma'}{\sum}V_{mn,\sigma\sigma'}^{l}c_{m+l,\sigma}^{\dagger}c_{n,\sigma'}^{\dagger}c_{n+l,\sigma'}c_{m,\sigma}\\
&+\underset{m\geq m_{0},\sigma}{\sum}(\triangle U_{mm_{0}}^{cp}+U_{mm_{0}}^{HF})c_{m,\sigma}^{\dagger}c_{m,\sigma}.
\end{split}
\label{effH}
\end{equation}
The first term is the same as the Coulomb interaction term in Eq. (3), except that here we only consider the edge electrons in orbitals $m\geq m_{0}$. The orbital $m_{0}$ is suitably chosen to study the edge instabilities if $\left\langle n_{m_{0}}\right\rangle \thickapprox 2$ in the exact diagonalization result and the choice of $m_{0}$ in our calculation will be stated below. The electrons in orbitals $m<m_{0}$ are defined as bulk electrons in our treatment. $m_{0}=(N-N_{edge})/2$, where $N_{edge}$ is the particle number of edge electron system we choose and $N$ is the particle number of the whole system.
$\triangle U_{mm_{0}}^{cp}\equiv U_{m}^{cp}-U_{m_{0}}^{cp}$, where $U_{m}^{cp}$ is the one body confining potential term defined in Sec. II. For positive background charge and linear confining potential $\triangle U_{mm_{0}}^{cp}$ is dependent of particle number $N$, while it is independent of $N$ for parabolic confining potential.
When edge instability happens, bulk orbitals are still completely occupied. So the Coulomb interaction between bulk and edge electrons has only Hartree and Fock terms left, which is equivalent to a one body Hartree-Fock field $U_{mm_{0}}^{HF}=\underset{m'<m_{0}}{\sum}(2V_{m'm}^{0}-V_{m'm}^{m-m'}) $ felt by edge electrons, and $V_{m'm}^{0}$ and $V_{m'm}^{m-m'}$ were defined in Eq. (4).

Actually the two assumptions at the beginning of this subsection for large electron systems are not very secure. In disk geometry the sharp peak of the wave function at the $mth$ orbital $\phi_{m}(r)$ is at $r=\sqrt{2m}l_{B}$, and the wave function has width $\sim l_{B}$. The larger the particle number, the  more electron orbitals are within the width $ l_{B}$ and may be involved in the instabilities in $\nu \lesssim 2$ regime.
For the phase transition from unpolarized state $\left|C_{N/2}^{N/2}\right\rangle$ to $S=4$ compact state $\left|C_{N/2+4}^{N/2-4}\right\rangle$ in our calculation, we choose the particle number of edge electron system as large as $N_{edge}=18$ (i.e. $m_{0}=N/2-9$) for $N \geq N_{edge}$, and give edge electrons $18$ orbitals. By this choice, the occupation number of the innermost edge orbital $\left\langle n_{m_{0}}\right\rangle$ is larger than $1.99994$; the occupation number of the outermost given orbital $\left\langle n_{m_{0}+17}\right\rangle$ is less than $0.00005$ for all our numerical results up to particle number $80$. Because bulk electrons do not have tendency to move outward, and larger edge orbitals have no tendency to be occupied, all the orbitals with occupation numbers not close to $0$ or $2$ are included in the edge electron systems we studied. Besides, we do not need to worry about the validity of the two assumptions for large particle number $N$ up to $80$, since for our choices of $m_{0}$ the exact diagonalization results and the assumptions are self-consistent.

\begin{figure}[t]
\includegraphics[width=8.5cm]{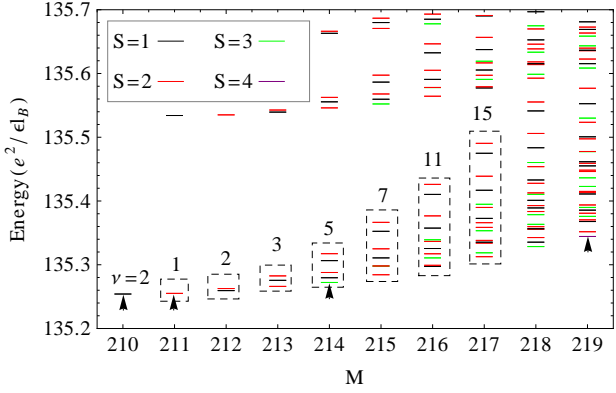}
\caption{(Color online) Low energy spectrum of $30$-electron system, showing the low energy excitations of $\nu=2$ unpolarized state under parabolic confinement when tuning parameter $b=0.14$ ($m_{0}=6$ in the calculation). The total spin quantum numbers $S$ of eigenstates are labeled by different colors as the annotation. The ground state is $\nu=2$ unpolarized state. The dashed boxes enclose the low energy excitations in each subspace close to the one of $\nu=2$ unpolarized state, and the amount of the states in each box is labeled above it. The states pointed by arrows are compact states.}\label{}
\end{figure}

In $\nu \lesssim 2$ regime, low energy spectra of electron systems with particle numbers up to $80$ under the three confinement are obtained by using exact diagonalization on the edge electron systems illustrated above. We will use the 30-electron system's low energy spectrum under parabolic confinement with $b=0.14$ ($\nu=2$ unpolarized state is the ground state) to show some common characteristics shared by these spectra. $\Delta M=M-M_{0}$ is the difference of orbital angular momentum quantum number $M$ compared to the one $M=M_{0}$ for $\nu=2$ unpolarized state ($M_{0}=210$ for $N=30$ system). In the subspaces with $\Delta M>0$, a set of low energy excitations (enclosed by dashed boxes in Fig. 12) are separated from higher energy states. By checking these low energy excitations' counting and spin configuration in each subspace $\Delta M$ with the ones predicted by $SU(2)$ effective theory,\cite{SU2effectiveTheory} we verified that these excitations are pure spin excitations and constitute the ESW branch of $\nu=2$ unpolarized state. One of these excitations becomes the ground state and thus destabilizes the $\nu=2$ unpolarized state with smoother confinement. Very interestingly, we observe that in each subspace $M$ which allows the existence of a compact state, the compact state (labeled by arrows in Fig. 12) {\em always} has the lowest energy. For this reason, in the following we view the compact states as the $2$nd generation descendants of the $\nu=2$ singlet state, and view some low lying non-compact states as their descendants. This provides us with an essentially complete understanding of the low-energy spectrum, and will make it an analogy of the spectrum of $\nu=1$ ferromagnetic state as stated in the following.

\begin{figure}[h]
\includegraphics[width=8.5cm]{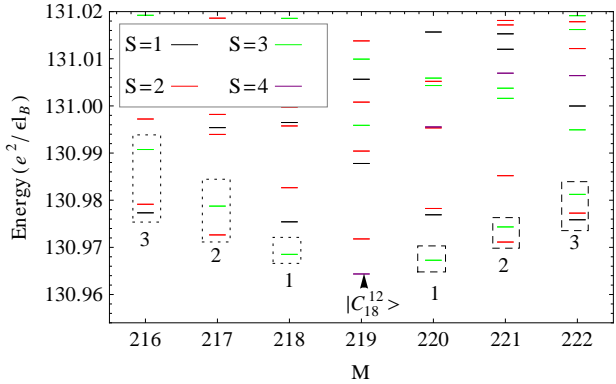}
\caption{(Color online) Low energy spectrum of 30-electron system, showing the low energy excitations on top of compact ground state $\left|C_{18}^{12}\right\rangle$ under parabolic confinement when tuning parameter $b=0.13$ ($m_{0}=6$ in the calculation). The different total spin quantum numbers $S$ of the eigenstates are labeled by different colors as the annotation. The ground state is compact state $\left|C_{18}^{12}\right\rangle$. The dashed boxes enclose the low energy excitations of compact ground state $\left|C_{18}^{12}\right\rangle$, and the amount of the states in each box is labeled below it.}\label{csspectrum}
\end{figure}

\begin{figure}[h]
\includegraphics[width=9cm]{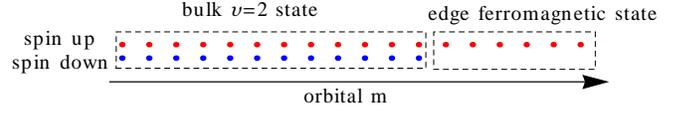}
\caption{(Color online) Sketch of the orbital configuration of compact state $\left|C_{18}^{12}\right\rangle$ with $S_{z}=S$. }\label{}
\end{figure}

\begin{figure*}[t]
\includegraphics[width=13cm]{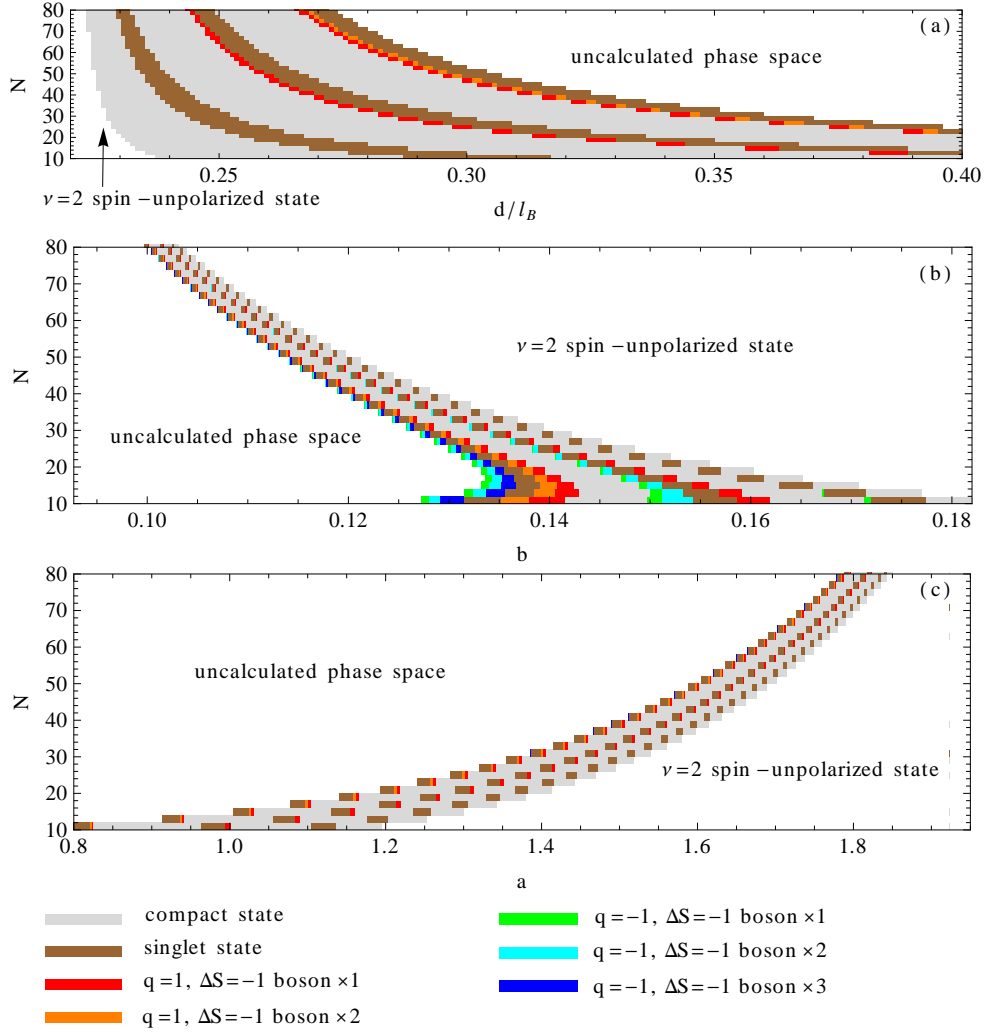}
\caption{(Color online) Phase diagrams in $\nu\lesssim 2$ regime for particle numbers from $10$ to $80$ (even numbers) under (a) $\nu=2$ positive background charge, (b) parabolic and (c) linear confining potential. Edge electron number $N_{edge}$ is chosen to $18$ for $N\geq18$, and $N_{edge}=N$ for $N<18$. The orbital number for edge electrons is $18$, so electrons can occupy from the $m_{0}$th to the $(m_{0}+17)$th orbital of the whole system. Step values are $\Delta (d/l_{B})=10^{-3}$, $\Delta b=10^{-5}$ and $\Delta a=10^{-5}$ for the three confinements respectively. Different kinds of states are labeled by the bars with different colors. In the annotation the bosons are the ones excited from the compact state in the stronger confinement side of the phase diagram if $q=1$ and the compact state in the smoother confinement side if $q=-1$. Singlet state can be view as the excited bosons on top of the compact states in both sides. }\label{v2phase}
\end{figure*}

In each subspace $M$, part of the low energy excitations can be mapped onto bosonic ESW excitations on top of a neighboring compact state, and just like in the $\nu=1$ ferromegnetic state, each boson reduces the compact state's total spin by $1$.
Taking the low energy spectrum of $30$-electron system under parabolic confinement around the subspace of compact state $\left|C_{18}^{12}\right\rangle$ with $M=M_{0}=219$ (Fig. {\ref{csspectrum}}) as an example, in subspaces with $\Delta M=M-M_{0}>0$ some of the low energy excitations are mapped onto the $\Delta S=-1$ bosons with positive angular momentum $q$ on top of compact state $\left|C_{18}^{12}\right\rangle$ (labeled by dashed boxes); in subspaces with $\Delta M<0$ some of the low energy excitations are mapped onto the $\Delta S=-1$ bosons with negative angular momentum $q$ on top of compact state $\left|C_{18}^{12}\right\rangle$ (labeled by dotted boxes). These compact state's bosonic excitations may mix with other low energy states in the same subspace, which are the excitations of other states in the compact state's subspace. The low energy excitations of a compact state are similar to those of $\nu=1$ ferromagnetic state, because both of them can be mapped onto the excited bosons which reduce the system's total spin by one on top of the original state. This is not surprising since a compact state can be viewed as a $\nu=2$ unpolarized state surrounded by a $\nu=1$ ferromagnetic state at the boundary as shown in Fig. 14. In $\Delta M>0$ subspaces, excitations of a compact state are determined by the excitations of bulk $\nu=2$ state and edge ferromagnetic state and also their correlation. In $\Delta M<0$ subspaces, the bulk $\nu=2$ state is inert and excitations of a compact state are only determined by the edge ferromagnetic state. In either case the ferromagnetic state at the edge of the compact state plays an essential role on the excitation.  By our exact diagonalization result, we found the excited bosons on top of the $S\neq 0$ compact states close to $\nu=2$ state have minimum at $q=1$ for positive $q$ and at $q=-1$ for negative $q$. Therefore, the instabilities of a compact state close to $\nu=2$ are just the condensation of $\Delta S=-1$, $q=1$ bosons (ESWs) at the smoother confinement side and the condensation of $\Delta S=-1$, $q=-1$ bosons (ESWs) at the stronger confinement side. Two neighboring compact states $\left|C_{N-k}^{k}\right\rangle$ and $\left|C_{N-k+1}^{k-1}\right\rangle$ can be connected by their excited bosons. As shown in Fig. 13, the lowest energy state in $M=222$ subspace with $S=0$ has three excited $q=1$, $\Delta S=-1$ bosons on top of compact state $\left|C_{18}^{12}\right\rangle$; while it can also be viewed having four excited $q=-1$, $\Delta S=-1$ bosons on top of compact state $\left|C_{19}^{11}\right\rangle$. The lowest energy excited boson of compact states close to $\nu=2$ having $q=\pm1$, explains our finding in last subsection that the ground states appear in between compact state $\left|C_{N-k}^{k}\right\rangle$ and $\left|C_{N-k+1}^{k-1}\right\rangle$ have their major charge and spin reconstructions on top of these compact states in a segment of orbitals which are only occupied by spin up electrons (the orbitals of edge ferromagnetic state) in state $\left|C_{N-k+1}^{k-1}\right\rangle$.

The phase diagrams in $\nu\lesssim 2$ regime are obtained from exact diagonalization for the particle number up to $80$ under the three types of confinement potentials in Fig. 15. They are qualitatively similar to those of small systems in Sec. IV.A. Compact states (gray regions in Fig. 15) with different total spins become ground states in some tuning paramter regions, and between two neighboring compact states, their bosonic excitations (ESWs) with $q=1$ or $-1$ may also destabilize them and appear as ground states. Singlet states which can be viewed as the condensation of maximum number of bosons from either of the nieghboring compact states which always appear as ground states (brown regions in Fig. 15). The bosonic excitations (ESWs) of compact states with $S \neq 0$ may or may not destabilize the compact states to become ground states. The windows of these compact states' ESWs in phase diagrams will change with different particle numbers, and different confinement types with the step values of our calculation.
Up to $80$ electrons, our exact diagonalization result shows that the initial $\nu=2$ instability is toward the $S=1$ compact state. As shown in Fig. 15, $\nu=2$ edge instability critical point under positive background charge confining potential approaches a constant $d_{c}/l_{B} \approx 0.223$, which indicates the existence of thermodynamic limit predicted by our electrostatic model. The $\nu=2$ instability critical point $b_{c}$ ($a_{c}$) for parabolic (linear) confinement decreases (increases) with increasing particle number, and these are also consistent with the prediction of the electrostatic model.

\begin{figure}[h]
\includegraphics[width=6.7cm]{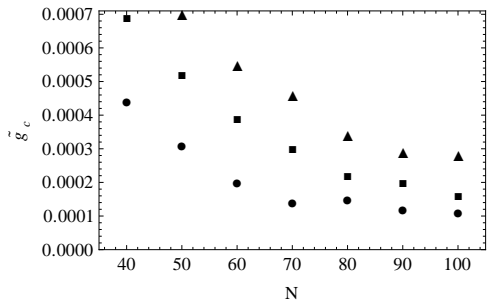}
\caption{Particle number $N$ {\em vs} renormalized critical $g$ factor $\tilde{g_{c}}=g_{c} \mu_{B} B/(e^{2}/\epsilon l_{B})$ in which the edge spin wave (ESW) phase window between neighboring compact states closes under positive background charge confinement. Circular points label the close of ESW phase window between $S=1$ and $S=2$ compact state; rectangular points label the close of ESW phase window between $S=2$ and $S=3$ compact state; triangular points label the close of ESW phase window between $S=3$ and $S=4$ compact state.}\label{z2}
\end{figure}
In the phase diagrams Figs. \ref{v2phase}, the ESW ground states have smaller spin quantum number $S$ compared to their two neighboring compact ground states. With finite Zeeman coupling, the ESW phase window will close and only compact states appear in the phase diagrams. We calculated the renormalized critical $g$ factors $\tilde{g_{c}}=g_{c} \mu_{B} B/(e^{2}/\epsilon l_{B})$ in which the ESW phase windows between neighboring compact states close for different particle numbers $N$ under positive background charge confinement (Fig. \ref{z2}). $\tilde{g_{c}}$ approaches a constant $\sim 10^{-4}$ with increasing particle number. Therefore we predict that in thermodynamic limit and under positive background charge confinement, with a small Zeeman coupling ($\tilde{g_{c}} \sim 10^{-4}$), the windows of ESW phase close and only compact states appear in the phase diagram near $\nu=2$.

\section{CONCLUding remarks}

In this paper, we investigate the low energy spin excitations and edge instabilities triggered by them in integer quantum Hall liquids, and the resultant edge reconstructions. We conclude that there are likely very rich spin structure for the edges of both $\nu=1$ and $\nu=2$ quantum Hall liquids. We also find that the specific form of the instability and the resultant edge structure is very sensitive to the details of the confining potential.

While we have the real electron spin in mind in this work, our study and likely many of our results can be generalized to systems with pseudospins, including bilayers quantum Hall systems\cite{girvinmacdonaldreview} and graphene.\cite{barlasreview12} In some sense these systems can be even more interesting. For examples there is no Zeeman coupling to the pseudospins and thus no associated energy penalty for pseudospin textures; the Dirac Landau level wave function allows for spin and pseudospin textures in higher Landau levels;\cite{alicea} and the lack of SU(2) symmetry allows more exotic forms of pseudospin texture in bilayer systems.\cite{moon}

Another natural direction to pursue is edge spin excitations and textures in fractional quantum Hall liquid. In fact a preliminary attempt has already been made in that direction.\cite{hu}

\section*{Acknowledgments}
We thank Zixiang Hu, Xin Wan, Yafis Barlas and Hua Chen for valuable discussions and assistance.
This work was supported by DOE grant No. DE-SC0002140.

\end{document}